\documentclass[12pt,preprint]{aastex}

\newcommand\etal{{\it et al.~}}

\newcommand\eg{{\it e.g.,~}}
\newcommand\ie{{\it i.e.,~}}

\slugcomment{Accepted, to appear in ApJ 10 November 2005, v633 2}

\begin{document}

\title{3D Simulations of MHD Jet Propagation Through \\ Uniform and Stratified External Environments}

\author{S. M. O'Neill\altaffilmark{1}, I. L. Tregillis\altaffilmark{2}, T. W. Jones\altaffilmark{1}, Dongsu Ryu\altaffilmark{3}}
\altaffiltext{1}{School of Physics and Astronomy, University of Minnesota, Minneapolis, MN 55455; smoneil@msi.umn.edu; twj@msi.umn.edu}
\altaffiltext{2}{Applied Physics Division, MS B259, Los Alamos National Laboratory, Los Alamos, NM, 87545; iant@lanl.gov}
\altaffiltext{3}{Department of Astronomy and Space Science, Chungnam National University, Daejeon, 305-764 Korea; ryu@canopus.cnu.ac.kr}

\begin{abstract}
We present a set of high-resolution 3D MHD simulations of steady light, supersonic jets, exploring the influence of jet Mach number and the ambient medium on jet propagation and energy deposition over long distances.  The results are compared to simple self-similar scaling relations for the morphological evolution of jet-driven structures and to previously published 2D simulations.  For this study we simulated the propagation of light jets with internal Mach numbers 3 and 12 to lengths exceeding 100 initial jet radii in both uniform and stratified atmospheres.
 
The propagating jets asymptotically deposit approximately half of
their energy flux as thermal energy in the ambient atmosphere, almost independent of jet Mach number or the external density gradient.  Nearly one-quarter of the jet total energy flux goes directly into dissipative heating of the ICM, supporting arguments for effective feedback from AGNs to cluster media.
The remaining energy resides primarily in the jet and cocoon structures.  Despite having different shock distributions and magnetic field features, global trends in energy flow are similar among the different models.
                                                                               
As expected the jets advance more rapidly through stratified atmospheres than uniform environments.  The asymptotic head velocity in King-type atmospheres shows little or no deceleration.  This contrasts with jets in uniform media with heads that are slowed as they propagate.  This suggests that the energy deposited by jets of a given length and power depends strongly on the structure of the ambient medium.  While our low-Mach jets are more easily disrupted, their cocoons obey evolutionary scaling relations similar to the high-Mach jets.

\end{abstract}


\section{Introduction}

The important roles of supersonic jets in various astrophysical contexts, such as radio galaxies, are now well-established.
Still, supersonic jet propagation and interaction with the ambient medium are complex, remaining topics of current research interest.
Studies of fully three-dimensional (3D) flows are quite limited, especially for jet propagation over distances giving length-to-radius ratios comparable to those observed.
Among several dynamical issues, two have particularly general importance; namely, the character of energy transfer from active jets to their environments and jet length and flow morphology evolution in time.
 
Energetic jets commonly are invoked as mechanisms responsible for disruption of cooling flows and heating of cluster environments (e.g. \citet{bohringeretal02,churazovetal02,blantonetal03,zannietal05}).
Once a jet ceases to be powered at its source, the energy in its cocoon may remain inside a buoyant bubble that does mostly adiabatic work on the environment until it is disrupted (e.g., \citet{bruggen02,fab03,robin04,jd05}).
At that time, the remaining bubble energy that has not been radiated or conducted away is shared. 
More immediate, critical questions concerning jets are how much of the energy flux carried by an active jet is shared with the environment and how much of that appears as irreversible heat.

Models of the evolution of flow morphology provide further insight into energy deposition by active jets.
Analytic models, such as those developed by \citet{cioffiblon92,kaiseralex97,komissarovfalle98}, and \citet{alexander00}, have been used to describe the expected morphology evolution of these systems in an attempt to understand their distribution in several observational planes.
Such simple self-similar relations allow for estimation of energy densities in these systems independent of detailed source histories.
Complimentary two-dimensional (2D) numerical simulations of jets, such as those done by \citet{normanetal82,coxetal91,falle91,cioffiblon92,hardeeetal92,komissarovfalle97}, and \citet{carvalhoodea02a, carvalhoodea02b}, have been conducted to examine the validity of these simple models. 
Simulations have illustrated the complexity of jet flows and have helped immensely in exploring the physical and environmental parameter space available to these objects.
Furthermore, \citet{zannietal05} recently described a set of 2D hydrodynamical simulations of jets propagating into realistic cluster environments.

Still, fully 3D simulations provide the best approach to addressing the issues raised above since such simulations can follow the highly complex flows driven by jets and allow a full accounting of the energy transport in these systems.
Previous 3D simulations such as those conducted by \citet{hardeeclarke92,hoodawiita96}, and \citet{norman96} have successfully illustrated flow features unique to three dimensions, the latter having simulated a 3D jet over 100 jet radii in length.
\citet{krause03,krause05} has further explored the propagation of 2D
axisymmetric and fully 3D evolved jets into uniform and King 
atmospheres, examining the shape of the bow shock structures over time and exploring the influence of boundary conditions on the propagation of jet-driven structures.

Our previous work, described in \citet{tregillisetal01a,tregillisetal01b,tregillisetal04} illustrated the complexity of the shock and magnetic field structures generated in full 3D jet-driven flows and described how detailed information about nonthermal particle populations in these systems is essential for correctly relating observations to the antecedent physical structures.
In the present work, we examine the long-term evolution of steady light three-dimensional magnetohydrodynamic (3D MHD) jets, exploring how the energetics, dynamics, and morphology of the bulk plasma evolve, and whether they do so in a simple manner.
To assure that fully 3D dynamics are as divorced as possible from startup
behaviors we follow the evolution of each jet and its
neighboring environment until the jets have penetrated
more than 100 initial jet radii into those environments.
We model both high and low Mach number jet behaviors and also consider uniform and stratified ambient media.
We examine for each system the time evolution of energy flow, 
computing the amount of inflowing energy deposited in the environment and determining how energy is partitioned within the disturbed flows.
We further observe how jet length and flow morphology evolve in time and compare our 3D results to those of 2D simulations and models to illustrate which features of jet-driven flows are well-described by simple models and which features differ in detailed treatments.

In $\S 2$, we discuss the details of our numerical methods and simulation properties.
Analysis of our simulated data is described in $\S 3$, while conclusions and astrophysical implications are discussed in $\S 4$.

\section{Calculation Details}
\subsection{Numerical Methods}
Our simulations are carried out on a 3D Cartesian grid. 
They employ a second-order total variation diminishing (TVD) nonrelativistic ideal MHD code, as described in \citet{ryuj95} and \citet{ryuetal98}.
The method conserves mass, momentum, and energy to machine accuracy.
To set up hydrostatic equilibrium in the ambient media external gravity is added as a source term in the $x$ direction through operator splitting, applying the $x$ momentum correction at each timestep to recalculate the total energy.
Preconditioning of the Riemann solver is included to maintain second order time accuracy. In this treatment of gravity, momentum and energy are no longer exactly conserved. 
However, we have confirmed that associated errors are much 
too small to influence our results.
The code maintains a divergence-free magnetic field at each time step using a constrained transport scheme (\citet{ryuetal98}).  
A gamma-law gas equation of state is assumed with $\gamma = 5/3$.  

A passive mass fraction or ``color'' tracer, $C_j$, is introduced at the jet orifice to track jet material as it propagates through the computational grid.  
$C_j$ is set to unity in the jet, while $C_j = 0$ in the ambient
medium.
Passive nonthermal, relativistic electrons are included, as well (see, \eg  \citet{jonesetal99,tregillisetal01a,tregillisetal04}) in order to model nonthermal emissions from the flows.  
We restrict our present analysis to study of the bulk flow, leaving the complimentary emission analysis to a separate paper (O'Neill \etal, in preparation).
     
Our simulated jets propagate approximately along the $x$ axis after entering the grid through a circular orifice centered in the $x=x_0=0$ plane.
The computational box extends to $x=x_1=230$ kpc, spanned by a grid of 576 uniform zones ($\Delta x \approx 0.4$ kpc). 
The equal, transverse, $y$ and $z$, dimensions of the box are selected
for each simulation so that they contain the entire jet bow wave until the end of the simulation.
Within 25 kpc of the box midline the transverse grid zones are uniform ($\Delta y = \Delta z = \Delta x$). 
Exterior $y$ and $z$ zone sizes expand logarithmically with an expansion factor 1.1, out to a maximum zone size of 8.4 kpc. Box dimensions for
each simulation are listed in Table 1.
 
Continuous boundary conditions are employed for both extremes of $y$ and $z$.
A modified continuous condition is applied at $x_1$, designed to maintain hydrostatic equilibrium in the undisturbed medium. 
Inflow boundaries are applied inside the jet orifice on the $x_0$ boundary. 
With the exception of one simulation (HU-r, described below) the same modified continuous boundary condition is applied on the rest of the $x_0$ plane as on the $x_1$ plane. 
In the HU-r simulation, reflecting boundaries are applied at $x_0$ outside the jet orifice.

\subsection{Simulation Properties}

We discuss five simulations, including Mach 12 jets (labeled `H' for `high Mach') and Mach 3 jets (labeled `L' for `low Mach').  
For each Mach number we simulate jets penetrating uniform media (labeled `U') and stratified, King-type media (labeled `K').
Except for the entrance plane of the jet ($x = 0$) all boundaries
remain undisturbed during all the simulations. 
This feature is necessary for us to examine the influence of the jet on the ambient medium and especially on its energetics. 
It is not possible to avoid influences from the
jet inflow plane, so long as that is a grid boundary.
To evaluate the role of this boundary we computed a pair of `HU' 
simulations; one with open (continuous) boundaries on this plane and one with reflecting boundaries (leading to the `HU-r' label). 
We briefly outline the properties of the jets and the ambient media
in the following two subsections.
The physical parameters of each simulation are summarized in Table 1.

\subsubsection{The Jets}
 
The jet inflows are steady after a brief starting sequence.
The incoming jet flow slowly wobbles in a 3 degree cone around the $x$ direction.
This establishes fully 3D flows within the physical domain as early as possible. 
Our five model jets are identical except for Mach number and period of the induced jet wobble.
The entering jets have uniform cores of radius $r_j = 2$ kpc surrounded by a concentric transition annulus that smoothly connects to the ambient conditions.
The jet core speed is $v_j = 0.15c$. 
The core density is $\rho_j = \eta \rho_0$, where $\rho_0$ is the ambient density at $x = 0$ (discussed below) and we set $\eta = 0.01$ in each simulation. 
The jets enter with the same gas pressure as the local ambient medium; namely $P_0$, as discussed below.
Earlier 2D studies have pointed to significant dynamical dependencies on jet Mach number (\eg \citet{carvalhoodea02a,carvalhoodea02b}). 
Our jets are parameterized by internal Mach numbers, $M_j$, making the internal jet sound speed, $c_j = v_j/M_j$.
We discuss simulations applying $M_j = 12$ and $M_j = 3$. 
As noted before, we designate the former as `high Mach' (`H') and the latter as `low Mach' (`L') jets.
The period of the inflowing jet wobble is 16 Myr for the `H' model jets and 4 Myr for the `L' model jets.

The inflowing jet power is calculated from $L_j \approx L_k+L_t$, where $k$ and $t$ refer to the kinetic and thermal components, respectively, ignoring the negligible contributions of magnetic and gravitational energy to the total power.  
The power components are given by $L_k=\int ((1/2) \rho_j v^2) v_x dA$ and $L_t=\int (\gamma(\gamma-1)^{-1} P_j) v_x dA$, where the integrals span the core and transition inflow regions.
Inserting physical values (see Table 1) and ignoring discretization
on the grid, this allows us to compute a nominal
jet power simply as a function of jet Mach number; 
namely, 
\begin{equation}
L_j \approx 1.12 \times10^{44}~(M_j^2 + 4.6)~{\rm erg~s}^{-1}.
\label{power}
\end{equation}
The actual, measured energy flow onto the grid is slightly less than this
estimate due to grid discretization and backreaction of the flows
on the grid near the perimeter of the jet orifice. The difference
is asymptotically only about 1\%, so this analytic expression
provides a very good measure of the energy added to the volumes being modeled (see \S 3.1).
To avoid start-up difficulties the jet speeds were ramped to full value over a finite time that depended on Mach number. 
Jet penetration and energy deposition during that time were negligible, 
so in the discussion below we reset the simulation clocks to start at the time when the jets reach full speed.

The jet magnetic field consists of toroidal and poloidal components.
The inflowing jet poloidal field equals the uniform ambient field, $B_{x0}$, discussed below. 
The toroidal field inside the jet core is $B_{\phi 0} = 1.25 B_{x0}(r/r_j)$.
This field component decreases quadratically to zero across the transition annulus.

Figures 1-5 show volume renderings of flow speed at late times for each of the five models.
Viewing these images and the associated animations provides
an efficient introduction to the behaviors of each simulation and
their intercomparison.
In each figure, the high-velocity jet enters the grid from the upper-right, inflating a cocoon of material that has entered the grid through the jet orifice, while the animations show the time evolution of these structures from several viewing angles.    
Propagation times across the grid vary with Mach number and especially with the structure of the ambient medium. 
They range from 26 Myr to 66 Myr, as listed in Table 1.

\subsubsection{The Ambient Medium}

We consider two simple model equilibrium atmospheres; namely a uniform (`U') and a plane stratified medium. 
The stratified ICM is a simple, isothermal King-type form \citep{king62} (`K'; see Figure 6),
\begin{equation}
\rho_a(x)=\frac{\rho_0}{[1+(\frac{x}{x_c})^2]},
\end{equation}
where $\rho_0$ is the density at $x_0$. 
For the K model atmosphere $x_c = \frac{1}{3} x_1 \approx 76.7$ kpc.
The U model atmosphere corresponds to $x_c \rightarrow \infty$.
The initial ICM pressure is simply $P(x) = c^2_a (\rho_a(x)/\gamma) = P_0 (\rho_a(x)/\rho_0)$, 
where $c_a$ is the ambient sound speed, and, for all simulations, $P_0= 1.43 \times 10^{-10}~{\rm dyne~cm}^{-2}$.
The ambient sound speed is set by the jet Mach number from the jet-ICM density contrast, $\eta$, and the assumption of jet-ICM pressure balance; namely, $c_a = v_j \sqrt{\eta} / M_j$. 
Accordingly the core jet speed has a Mach number with respect to $c_a$ given by $M = M_j/\sqrt{\eta}$. 
For our jet parameters the associated ICM temperatures for the $M_j = 12~(3)$ jets are $0.88~(14)~\mu$ keV, where $\mu$ is the mean molecular weight of the ICM.
These parameters are chosen such that the `H' models include temperatures appropriate for cooling-flow clusters with cooled cores while the `L' models describe hotter massive clusters.
The base ICM density is then, $\rho_0 = \gamma P_0/c^2_a = 1.18\times 10^{-29}~M^2~{\rm g~cm}^{-3}$.
The inferred gravitational acceleration is 
\begin{equation}
g(x)=-\frac{2 c_a^2}{\gamma}\frac{x}{x_c^2+x^2},
\end{equation} 
which vanishes in the `U' models. This gravity model is not truly
representative of those in real clusters. Its only purpose here is
to allow establishment of hydrostatic equilibrium in the
undisturbed stratified medium. Otherwise gravity plays a
negligible role in these simulations.
The initial ambient magnetic field in each simulation is uniform and in the $x$ direction; i.e., $B_{x} = B_{x_0} = 3~\mu$G.
The resulting magnetic pressure at $x_0$ is 1\% of the gas pressure; i.e., $\beta_0 = P_g/P_B = 100$. At the top of the King atmosphere $\beta = 10$.
This value of $B_{x0}$ is in the range of values suggested by observations of cluster media. 
The specific value was selected to control the synchrotron loss times of the relativistic electrons transported in the simulated flows.
While that issue is not included in the present discussion, it will be important to our subsequent discussion of the radiant luminosity evolution of the simulated objects (\citet{oneilletalprep}).
 
\section{Discussion}

\subsection{Energy Flows}

By isolating contributions from jet plasma and the ICM using the ``jet color'' tracer, we can characterize where energy in the system is transported and how much of it goes toward direct heating of the ambient medium.
Quantitatively following the flow of kinetic, thermal, magnetic and gravitational energy in each simulation further allows us to characterize how energy in the system changes form.
We begin our analysis by examining how energy brought onto the grid
by the jet is exchanged with the ambient medium and how it becomes
partitioned among different energy forms as a result of the complex
dynamics of the jet-ICM interaction. The generation of thermal
energy in the ICM and especially dissipated heat is
of special significance.
We also examine how this energy flow is affected by the structure of 
the ambient medium and jet Mach number.
Since the final propagation lengths of all our jets are the same, but
the propagation times span a broad range, it is most convenient to
present many of our results in terms of jet length rather than time. In
the following subsection we explore the dynamical time evolution of
each jet, including its length evolution, so that
one can translate length behaviors into
time behaviors, if desired (see \S3.2 and the upper two panels of Figure 18). In these discussions
we define the length of the jet as the largest value of $x$
contained by the bow shock preceding the jet. Typically
the jet (beam) terminus and the extremum of the bow shock are
at almost the same location. That position also
defines the `head' of the jet, $x_h$, so the velocity of the jet head
refers to $d x_h/dt$. We note, however, that the head of the jet
is not generally on the axis defined by the jet
orifice, due to the jet wobble and especially the sometimes dramatic dynamical
instabilities experienced by the jet tip.

The lower right panels of Figures 7-11 provide a basic accounting of
the energy changes introduced on the grid in each simulation. In
each case the solid line represents an integration of the jet power, given approximately by equation (\ref{power}), while
the dotted lines represent the measured total change in energy since the simulation began. Figure 7
illustrates the result for the HU-r model, where no energy
is allowed to leave the grid. Ideally we would expect the
two curves to overlap in this case. As noted above, they agree well, but not exactly,
the measured increase being slightly smaller. The difference comes from the
backreaction of the jet cocoon on the jet perimeter near the
orifice. That effect diminishes with time, so that asymptotically,
the nominal and actual energy fluxes agree to within about 1\%.

All the other simulations utilize open or continuous boundaries
along the jet inflow plane. As illustrated in Figures 8-11,
and as one would expect, the measured energy changes in
those cases are reduced by outflows across this plane. Still,
those losses are quite modest, being asymptotically
$\lesssim 13\%$ for all models.
We defer to \S 3.2 a discussion of the dynamical influences of
$x=0$ boundary conditions.

Having established a reasonable accounting of the global energy changes
in the different simulations we next examine the energy transferred
to the ambient medium from the jet penetrating it. To do that we
isolate the jet and its cocoon using the color tracer, $C_j$.
Figure 12 shows as a function of jet length the fractional change in kinetic and thermal energy in the ambient medium compared to the total, measured energy added to the grid.
These fractions are found by integrating each energy form over the
computational grid, weighted by the passive color tracer (actually $1-C_j$).
We note that this result differs by at most a few percent from
that obtained by isolating and removing the jet/jet-cocoon using a color threshold,
such as $C_j = 0.9$. Thus, relatively little energy has been
exchanged by entrainment (indicated by intermediate values of $C_j$)
during the periods simulated.
The plots show that in each of the `H' models approximately 55-60\% of the
jet energy is transferred
to the ambient medium; while about 45\% of the `L' model jet energies
are given to the ambient medium. Similarly
approximately 40-45\% of the jet energy is converted
in each case to ambient thermal energy. Those transfer fractions,
and especially the total energy measures,
are roughly constant once each jet has penetrated more
than about 50 jet radii into their environments. Thus, these
figures seem to represent fair estimates of the steady, asymptotic
energy exchange rates between such jets and their surroundings.
It is remarkable that they depend only weakly on the
Mach number of the jet or on the density profile of the
ambient medium.
Additionally, we estimate that between 40-60\% of the thermal energy entering the ambient medium is added irreversibly, mostly through shock dissipation.
The ambient irreversible (entropy enhancing) energy change estimate is 
obtained by subtracting adiabatic changes due to compression, given by $\Delta E_{ad}=-\int dV \int \gamma (\gamma-1)^{-1}P({\bf \nabla} \cdot {\bf v})dt$, from the measured change in ambient thermal energy.
Thus, a net  $\sim 20-30$\% of the steady jet power goes directly into entropy generation
within the ICM.
The efficient transfer of thermal energy from our simulated jets to their environments is significant, because it gauges the potential for active galaxies to provide energetic feedback to cluster environments while their
jets are `on'.  

\citet{zannietal05} have recently examined this same question
based on 2D, axisymmetric jet simulations. Their jets were
switched off relatively early, before they had penetrated more
than about 20 jet radii into their environments. So, close
comparison of our results with theirs is difficult. However, from their
Figure 8 we can estimate that in each of several
simulations about 15\% of their jet energy was
dissipated irreversibly at the time the jet was switched off. That is very
consistent with our findings. Their simulated jet parameters
extended to higher Mach numbers and larger density contrasts than ours.
However, once again, their dissipation estimates were relatively
insensitive to the jet parameters, reinforcing our
conclusion above about the general nature of this result. 
Subsequent to jet switch-off,
as one would expect, the bow shock of their inflated jet bubbles continued to dissipate an 
increasing fraction of the injected energy. In their simulations
$\sim 60-70$\% of the injected energy was eventually dissipated during that phase,
consistent with the classical dissipation behavior of a spherical
blast wave \citep{sedov59}.
MHD simulations of
subsonically inflated buoyant bubbles indicate that once the
bubble expansion is no longer supersonic relatively little
entropy is added to the ambient medium except through mixing \citep{jd05}.

Examining now in more detail global energetics, we
explore how the injected energy is partitioned over the entire grid,
including the jet/jet-cocoon and the ambient medium.
Figures 7-11 show how the partitioning of energy evolves as a function of total jet length for each model.
In each plot, the nominal energy input from the jet as defined by
equation (\ref{power}) is shown as a solid line for reference.  
The dashed lines in each plot indicate the nominal energy
input of the particular form (\eg kinetic, thermal, or magnetic). 
Dotted lines reveal the energy increment of each form actually
measured on the grid.
The vertical ordering of the dashed and dotted lines indicates the 
direction of energy conversion 
(\eg {\em dashed lines} above {\em dotted} imply that less energy of 
that type is measured than was added), while the distance between 
these lines reflects the amount of energy being transferred to/from a particular form.  
The lower right plot in each figure provides a comparison between the 
nominal total energy influx and the measured change in total energy on the grid for that model.
As mentioned previously, except for model HU-r, all
the models leak some energy across the jet inflow plane, $x=0$,
due primarily to escape of cocoon backflow.
Those losses are generally less than 13\% of the inflow energy, however, by the time jet propagation
is well established (\ie once the jet head is well away from
the $x=0$ boundary). By that time backflow velocities near the base of the cocoon as measured in the lab frame
are generally only a few percent of the jet
speed, while the densities and pressures are much reduced from those
in the cocoon head. Consequently, despite the large cross section of
the cocoon base, energy fluxes across the boundary are modest.
The leakage initially makes up a larger fraction of the inflow energy in the `L' models, but the asymptotic behaviors of the `L' and `H' models are similar.
In any case, the energy leakage in the open boundary
cases is too small to influence our conclusions.

The global energy flow characteristics of all five simulations are
qualitatively similar, especially for cases with the same
Mach number.  
In each case most of the inflowing jet energy is kinetic, of
course, because the jets are supersonic and superalfv\'enic.
The measured kinetic energy increase, however, is always 
significantly less than the nominal
kinetic energy provided by the inflowing jet, indicating that 
there is always a net 
conversion of jet kinetic energy into other forms. Gravitational
energy is a minor component in all the simulations, and it does not
become a significant reservoir. Changes in gravitational
energy are generally less than 1\% of the total energy added.
As mentioned previously, gravity is included in these simulations only as a method of stabilizing the undisturbed ambient medium, so we ignore it in our discussions.
Simulations such as those conducted by \citet{zannietal05} and \citet{krause05} have shown that gravitational energy changes can play a significant role in realistic cluster potentials, but we did not intend to investigate this issue with our models.
Although magnetic energy also remains small, it does increase
in all the simulations by a substantial factor. Since those changes
may be reflected in observable nonthermal emission properties, we will
include some comments below about magnetic energy variations.
Most of the transformed kinetic energy reappears as thermal energy, as our
discussion starting this section would suggest. There we emphasized
that the fractional energy transferred from the jets to ambient thermal
energy was similar for all the models during their asymptotic evolution.
That similarity applies also to the global transfer of thermal 
energy.

The relative model independence of the fractional thermal
energy transfer is remarkable partly because a large portion of the
thermal enhancements is due to shocks, while shock distributions
are rather different among the models. 
To illustrate this last point graphically, Figures 13 and 14 (and the associated animations) show volume renderings of the flow compression rates ($-{\bf \nabla}\cdot {\bf v}$) for the HU and HK runs.
Compression rate is a convenient shock tracer
that also provides a qualitative indicator of shock strength.
In Figure 13 (HU), we see a complicated 'shock-web complex' similar to distributions seen in earlier jet simulations in uniform media
(e.g. \citep{jonesetal01}). The shock web is most developed in the
jet backflow cocoon, where it is produced by flailing of the jet terminus
and especially by collisions between the jet and the wall of its cocoon
\citep{jonesetal01, coxetal91}.
Some of those shocks are moderately strong, as pointed out previously for
similar simulations \citep{tregillisetal01b, tregillisetal04} although many are weak.
The same jet-cocoon interactions also generate weaker shocks that penetrate into the
ambient medium inside the bow shock. All
of these shocks contribute to dissipation of jet kinetic energy.
In contrast to the HU case, the HK jet illustrated in Figure 14 
exhibits a noticeably more stable
propagation, and only near the very end of the simulation does
it begin to develop an evident shock web near its head. 
This difference in jet behaviors in uniform and stratified
media derives from the strong deceleration of the jet
head that develops in uniform media in contrast to an almost
constant extension of the jet in the `K' model environments.
The former leads to a much stronger backreaction on the
jet tip  from the ambient medium.
The similar fractions of jet energy that are irreversibly
dissipated under these different circumstances
emphasizes that in either case the jet thrust must
be transferred to its surroundings, which roughly defines the
pressure surrounding the jet as it propagates. That pressure
is produced largely through shocks, which can be simple and strong or 
complex but individually weak, so long as their accumulated effect is the same.

Just as we have seen for the thermal energy generation,
Figures 7-11 show that the total magnetic energy increases exceed
in each model the magnetic energy represented by the jet Poynting flux.
Early in the simulations this excess is typically a factor of two or so
in each case. Except for the LU simulation the relative magnetic
energy enhancement increases to at least a factor of five by
the end of each simulation. Just as it was for shocks, the
distributions of magnetic fields are rather different, despite the
similar energetics. This is illustrated by the contrasting
magnetic field distributions in the HU and HK simulations
shown in Figures 15 and 16 (and the associated animations). Those differences derive from the
same dynamical distinctions mentioned in reference to shocks.
Most of the magnetic field enhancements can be traced to compression
events in the flows (especially shocks), followed by flux
stretching. 

We end this section by pointing out one important detail that results from
the differences in propagation histories of the `U' and `K' simulations
for a given power class (`H' or `L').
In particular, since jets propagating through a stratified atmosphere
exhibit much less deceleration, their lengths are greater for
a given age. Consequently, if we consider the total
energy budget of jets of a given length, the energy
deposited by the `U' jets will be substantially greater.
For example, by the time the jets have reached $90\%$ of the grid 
length, the HU jet has advected onto the grid twice as much energy as 
the HK while the LU has advected 1.75 times as much energy as the LK.
For both sets of Mach numbers, this difference is appreciable and is noticeable in the backflow, especially in the high-Mach jets.
Figures 2 and 3, for example, show that there is more material flowing opposite the direction of jet propagation in the HU case than in the HK, which is indicative of a greater amount of kinetic energy in the system at a given jet length. 

\subsection{Dynamics and Morphology}

We now explore some basics of the dynamical evolution of our 
simulations, with emphasis on how the cocoon length and volume
depend on the ambient density structure. We pay particular attention to asymptotic behaviors,
comparing them to simple models and previous 2D simulations.
The most basic dynamical features of jet evolution are the length, $l$ and 
the speed of the jet head, $v$, as functions of time.  
The simplest approach to modeling these uses dimensional
analysis based on the energy deposited by the jet; \ie one
applies an extension of the familiar Sedov-Taylor analysis
(\eg \citet{falle91,heinz98, zannietal05}).
Our simulated jets are steady, and we found above that a relatively
constant fraction of the jet power is transferred into thermal
energy in the cocoon and the shocked ambient medium, at least
asymptotically. We assume, therefore, that the energy
being stored increases linearly with time. 
Similarly, the ratio of energy in the shocked
ambient medium and the cocoon becomes relatively steady. 
The more directly observable of these volumes is likely
to be the cocoon, since it corresponds to the radio lobe of a radio
galaxy. Thus, we use that volume for our dimensional analysis and
assume that the energy content of the cocoon is a proportional to the
energy shared with swept up ambient matter.
To estimate analytically the mass swept up by the cocoon
one must make some assumption about the cocoon volume as a function
of length. It is common to assume a sphere, although to obtain a
scaling it is really necessary only to assume a volume form, $V(l)$.
A convenient  generalization  of the spherical assumption
is $V(l) \propto l^{1+\alpha}$. Any homologous structure, including
the sphere would correspond to $\alpha = 2$. 
Assuming the ambient density follows a simple power law of the form $\rho = \rho_0(
x/x_c)^{-\kappa}$, the {\it similarity length}
then scales as $l \propto t^{3/(3+\alpha-\kappa)}$, while the
{\it similarity speed} follows the form 
$v \propto t^{(\kappa-\alpha)/(3+\alpha-\kappa)}$, or $v \propto l^{(\kappa - \alpha)/3}$.

Figure 17 plots the measured cocoon volumes for each of our 
simulated flows as functions of jet length, while Figure 18
shows the measured length evolution and the measured head propagation
speed for each case. 
The cocoon volume corresponds to material with $C_j \ge 0.90$.
Although this method of accounting includes the jet in the cocoon volume estimate, the volume of the jet itself is always less than $10\%$ that of the cocoon, and this fraction decreases in time after the flow is well-established.  
It is worth noting that the characteristic cocoon widths, as estimated from assuming a cylindrical cocoon of the observed volume and length, are comparable in magnitude to the projected amplitude of the jet wobble, differing by roughly a factor of two at most.
We did not explore the dependence of this characteristic cocoon width on jet opening angle, but we do note that a cocoon volume dominated exclusively by the initial jet wobble would produce $V \propto l^3$ at all times, which is not seen.

The early cocoon evolution, when $l \lesssim 20-30$ kpc,
seems to follow a form close to $V \propto l^{3/2}$, or $\alpha = 1/2$,
for all the simulations. This interval corresponds roughly to
the times before the jets are recollimated after entering the grid.
All the jets are still well inside uniform ambient media
on those scales, so the anticipated similarity forms would
be $l \propto t^{6/7}$; $v \propto l^{-1/6}$.  The early 
velocity and especially the length evolution poorly approximate this,
showing, instead a flatter relation.
This breakdown of self-similar evolution should not be too
surprising in the initial dynamical stages, partly on account
of residual start-up behaviors. The energy partitioning
is not steady during this interval. 
This is also the regime where the simulations
with open boundaries along the $x = 0$ plane suffer the most
significant energy losses. We note, however, that the HU-r simulation,
which does not suffer such losses, is not a particularly closer
match to the self-similar form.
We should keep in mind, as well, that the flow pattern is still only a 
few jet radii in size (recall that $r_j = 2$ kpc). 

On intermediate scales, roughly $20-30~{\rm kpc}\lesssim l \lesssim 70~{\rm kpc}$,
all of the volume behaviors are consistent approximately with
$\alpha \sim 1/3$. This regime corresponds to the intervals between
jet recollimation and when the `K' model jets begin to emerge from
the core density region. In this regime the volume scaling
shows that the length of
the jet extends much more rapidly than the width of the cocoon.
Looking again at Figure 12 we see that the thermalization of
jet power still has not reached its asymptotic level. The cocoon is
not being driven strongly laterally by internal pressure.
Again setting $\kappa = 0$ in this domain,
the anticipated similarity length and velocity scalings would be 
$l \propto t ^{9/10}$, $v \propto l^{-1/9}$.
Here we find in Figure 18 a much better match with the measured length and velocity
behaviors for all of the jets.

Eventually the volume scalings steepen into something approximating
$V \propto l^3$ ($\alpha = 2$), but only after the jet lengths exceed about 50 $r_j$
(100 kpc). This transition corresponds to the cocoons becoming laterally driven strongly by internal pressure.
A fiducial line with slope 3 is included in Figure 17
for comparison. Since the `K' model jets extend into the strongly
stratified medium by that time, two different scaling relations
are expected. For the `U' media, with $\kappa = 0$ the forms are
$l \propto t^{3/5}$; $v \propto l^{-2/3}$. For the `K' models,
$\kappa = 2$ here, so we expect $l \propto t$; $v = {\rm constant}$.
Looking at Figure 18 and using the fiducial lines provided we
can see a fairly good correspondence with the measured behaviors
once the jet lengths exceed $\sim 100$ kpc.
We conclude, therefore, that asymptotically our jets and their
cocoons do evolve consistent with simple self-similar forms.
These behaviors are roughly consistent with those found from 2D
axisymmetric models and similar dimensional arguments by 
\citet{carvalhoodea02a,carvalhoodea02b}.  
We note that a more realistic model of cluster gravity and a full 3D spherically symmetric King profile has the potential to affect evolved flow structures by squeezing the cocoon, as discussed by \citet{alexander02} and seen in simulations by \citet{krause05}.

We add that \citet{carvalhoodea02a,carvalhoodea02b} also modeled explicitly
the evolution of
the cocoon volume based on their 2D axisymmetric simulations. 
Their 2D results for 'K' model atmospheres are comparable to our 'K' model
atmospheres in three dimensions. The scaling expectations, where in all cases $\alpha \approx 2$, thus are consistent with both the 2D results and our asymptotic 'K' models.
The uniform atmospheres in \citet{carvalhoodea02a} produce $\alpha \approx 2/3-7/3$ for the HU parameters and $\alpha \approx 5/4-3/2$ for the LU parameters.
For the HU run, this range includes our value of $\alpha \approx 2$, but our LU cocoon evolves slightly outside of the range of the 2D version.
As in the case of jet length, the evolution of our LU model more closely resembles that of our HU model than that of the 2D analog, indicating that, for our range of Mach numbers, 3D morphology evolution is not strongly influenced by Mach number.
Although the flows may have distinct morphologies from one another, the evolved jet lengths and cocoon sizes are all adequately described by simple scaling relations listed above.
  
To conclude this section we briefly compare the dynamical evolution of the
HU and HU-r simulations. Recall that the only difference between these
was the character of the boundary along the jet
inflow plane, $x = 0$. The HU simulation left this boundary open outside
the jet orifice,
while the HU-r simulation applied reflection conditions there, in order to
prevent any energy or mass loss there. As Figures 17 and 18 illustrate,
the asymptotic volumes, lengths and head speeds of these two simulations
are very similar, so that the choice of boundary condition has
not played a major role in their long term evolution. 
The HU jet is slightly longer at a given time, while
at the end of the simulations the HU-r cocoon volume is slightly larger for
a given length. At intermediate times and lengths, however, the
evolutionary behaviors are more obviously distinct. In particular,
the HU-r cocoon inflates significantly faster, while the HU head
advances significantly faster. Comparison between Figure 7 and Figure 8
reveals that the thermal energy enhancement is still significantly
greater in the HU-r simulation during this interval. The open boundary
in the HU simulation has reduced the accumulated cocoon energy near
the inflow plane, reducing the cocoon inflation there. As the jet
length increases, however, pressure gradients close to the
inflow plane become smaller, so that less and less of the
jet power escapes across this plane, and the boundary plays a smaller
and smaller role in the evolution of the jet and its cocoon. 

\section{Conclusions and Astrophysical Implications}
We have conducted a series of five simulations to explore the 
influence of Mach number and external environment on the large-scale 
propagation of steady, light MHD jets. 
We explored the flow of energy in our simulated systems, tracking both the global partitioning of energy in its various forms and the spatial distribution of energy as the jets evolved, seeking to determine the extent to which energy is transferred from these jets to their surroundings.
We further examined the morphologies and dynamics of our simulated jets and compared our data to simple dimensional scaling expectations and the results of analogous simulations in two dimensions.

Our simulations suggest that energy transfer from evolved jets to their environments is remarkably efficient.  
Approximately half of the energy flux of an evolved jet enters the ambient medium, and roughly half of this (one-quarter of the total flux) goes directly toward dissipative heating of the ambient medium.
This trend is present for all of our models with various Mach numbers and ambient medium structures, suggesting that energy transfer and dissipation is not strongly dependent upon the exact nature of the flow.
This result is important because high rates of energy flux into the ambient medium make AGNs good candidates for reheating of their ICM environments.

We also find that simple energy-based similarity scaling laws reasonably describe the asymptotic
time evolution of jet length and cocoon size in our simulations, independent of our model parameters.
This lends further support to the employment of these simple relations in estimating radio lobe energy densities and estimating flow properties.
Although some observed structures may depend upon the unique history of a particular object, our simulations suggest that very basic size parameters are well-described by simple scaling laws.

To reiterate, the most important results from our work are: 

1. Energy transfer to the ambient medium is efficient for all of our simulations, with approximately half of the inflowing jet energy becoming thermal energy in the ambient medium while the jet propagates. Roughly half of this energy is added irreversibly through dissipation, mostly at shocks.  
This characteristic behavior is independent of our models, suggesting that active jet reheating of the ICM can be efficient under a variety of physical circumstances, as is the case in 2D simulations.
Furthermore, this trend is independent of the differences in shock and magnetic field structure that occur in our various models.
This suggests that real radio galaxies may transfer energy equally efficiently to the ambient medium, despite having a variety of brightness distributions.

2. Dynamically, our jets asymptotically resemble simple dimensional scaling expectations and previous results in 2D.
This should suggest that supersonic jets are well-described by simple models for their behavior, and that it is reasonable to use such
scalings as starting points for models of jet luminosity and morphology evolution.

3. Since jets propagate to greater lengths in a given time through
stratified media than through uniform media, 
the jet lobe energy contents from jets of fixed power
and length depend significantly on the density profiles of their
environments. 
For a given Mach number and jet length, this has the effect of introducing more energy into uniform environments than stratified atmospheres.
Additionally, as our jets are disrupted, they convert a larger fraction of their inflow to thermal energy.
This implies that stalled jets in uniform atmospheres may eventually become even more efficient at generating thermal energy, most of which will eventually enter the ambient medium.

\acknowledgments
This work by SMO and TWJ was supported by the NSF grant AST03-07600 and by the Minnesota Supercomputing Institute.
The work by DR was supported by the KOSEF grant R01-2004-000-10005-0.  
Also, we wish to thank Brian Cornell for his work in developing some of our visualization tools.

\clearpage

\begin{deluxetable}{crrrrrrr}
\tabletypesize{\scriptsize} 
\tablecaption{Summary of Simulations}
\tablewidth{0pt} 
\tablehead{
\colhead{ID \tablenotemark{1}} & \colhead{$M$ \tablenotemark{2}} & \colhead{$M_j$} & \colhead{Atmosphere} & \colhead{$x_{size}$} & \colhead{$y,z_{size}$\tablenotemark{3}} & \colhead{Final Age} & \colhead{$L_j$ \tablenotemark{4}}}
\startdata
HU & 120 & 12 & Uniform (U) & 230 kpc & 228 kpc & $\approx 52$ Myr & $1.67 \times 10^{46}~{\rm erg~s}^{-1}$\\ 
HU-r & 120 & 12 & Uniform (U) & 230 kpc & 633 kpc & $\approx 52$ Myr & $1.67 \times 10^{46}~{\rm erg~s}^{-1}$\\
HK & 120 & 12 & King-type (K) & 230 kpc & 228 kpc & $\approx 26$ Myr & $1.67 \times 10^{46}~{\rm erg~s}^{-1}$\\
LU & 30 & 3 & Uniform (U) & 230 kpc & 903 kpc & $\approx 66$ Myr & $1.53 \times 10^{45}~{\rm erg~s}^{-1}$\\
LK & 30 & 3 & King-type (K) & 230 kpc & 633 kpc & $\approx 37$ Myr & $1.53 \times 10^{45}~{\rm erg~s}^{-1}$\\
\enddata
\tablenotetext{1}{HU-r model features reflecting boundaries surrounding $x=0$ jet orifice.  All other models have open boundaries everywhere.}
\tablenotetext{2}{Jet speed fixed at $v_j = 0.15c$.  Mach number adjusted with ICM density and sound speed.}
\tablenotetext{3}{Grid size designed always to include bow wave.  The outer zones in $y$ and $z$ dimensions are never disturbed, and the total grid size in some cases greatly exceeds the size of the disturbed regions.}

\tablenotetext{4}{Jet densities calculated from fixed $\rho_{j} = \eta~\rho_0$, with $\eta = 0.01$. ICM base density given by $\rho_0 = 1.18\times 10^{-29}~M^2~{\rm g~cm}^{-3}$.  Along with the ICM base pressure $P_0=P_j=1.43 \times 10^{-10}~{\rm dyne~cm}^{-2}$, this gives the ICM sound speed $c_a = 4.50 \times 10^9~M^{-1}~{\rm cm~s}^{-1}$. Jet magnetic fields include a poloidal
component equal to the ambient field $\beta = 100$ at $x = 0$ as
well as a toroidal component whose peak value is twice the
poloidal value.}
\end{deluxetable}

\clearpage

\begin{figure}
\epsscale{0.5}
\plotone{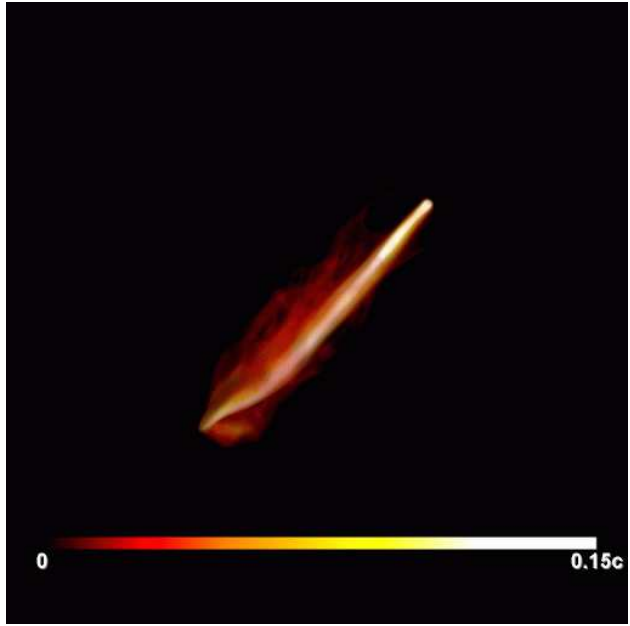}
\caption{Volume rendering of flow speed for the HU-r model, after the jet has propagated $\sim 90\%$ of the total grid length.  An animation of this quantity as seen from several different angles appears at: http://www.msi.umn.edu/Projects/twj/newsite/projects/radiojets/movies/ }
\end{figure}

\begin{figure}
\epsscale{0.5}
\plotone{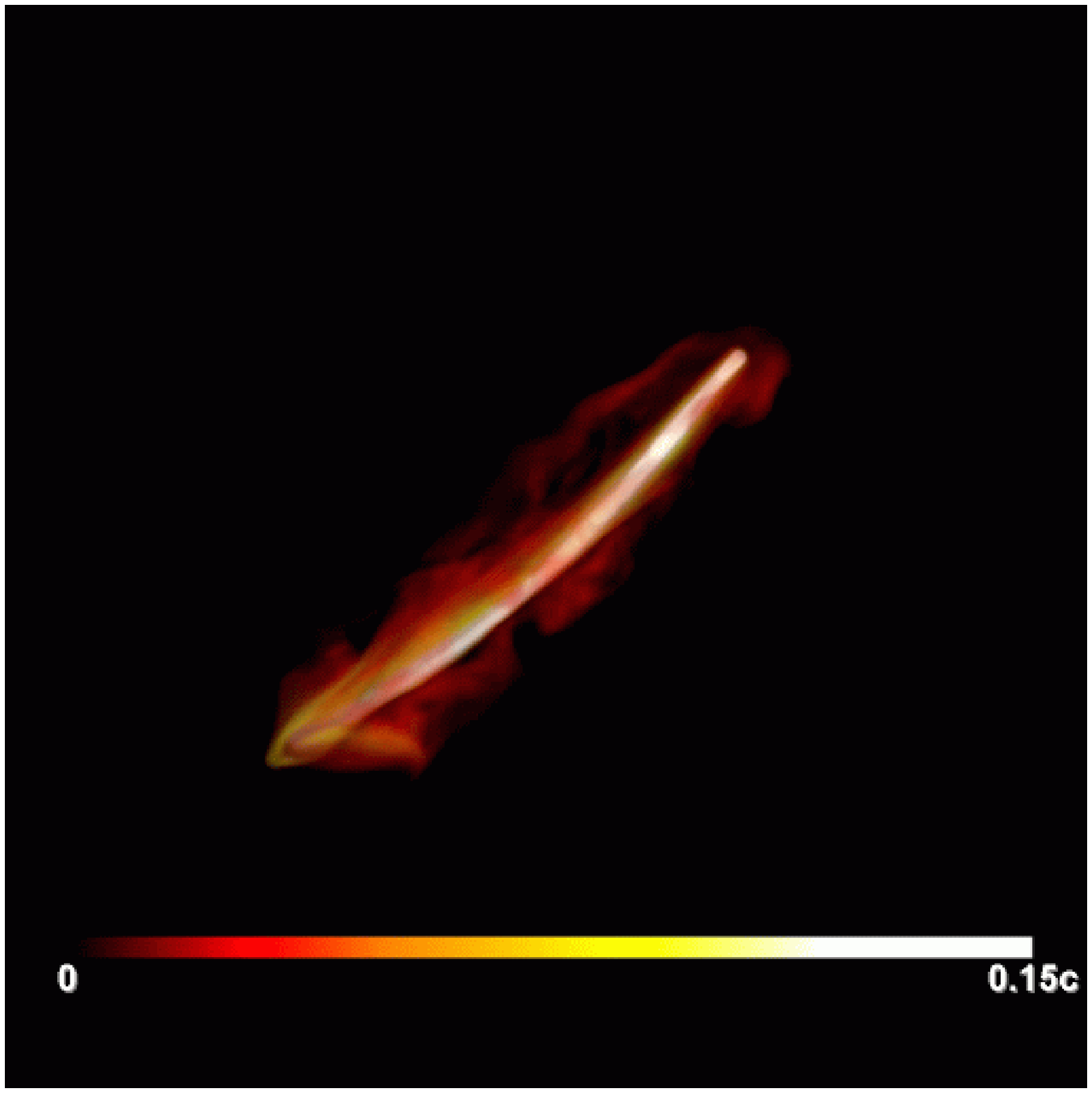}
\caption{As in Figure 1, for the HU model.}
\end{figure}

\begin{figure}
\epsscale{0.5}
\plotone{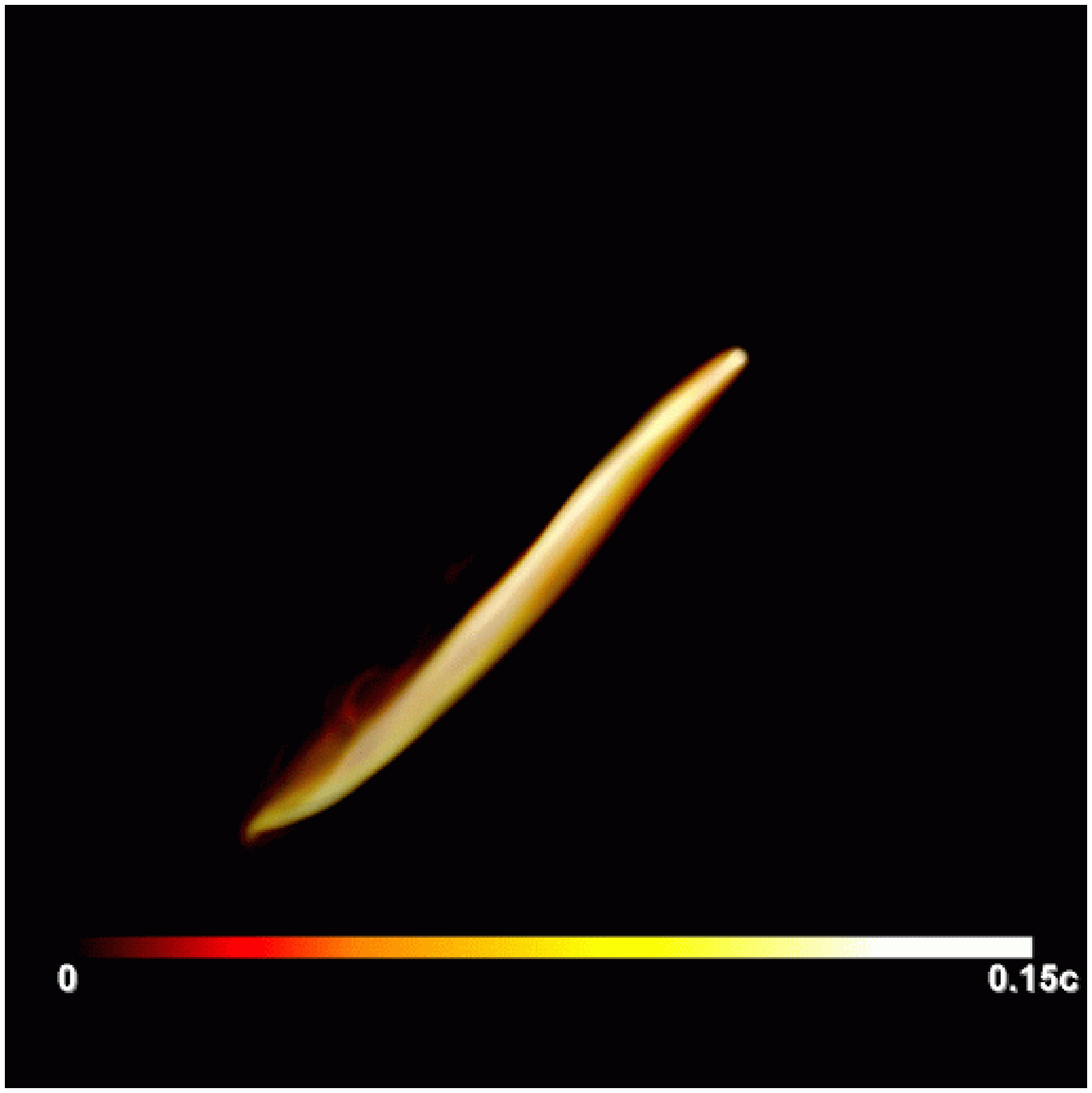}
\caption{As in Figure 1, for the HK model.}
\end{figure}

\begin{figure}
\epsscale{0.5}
\plotone{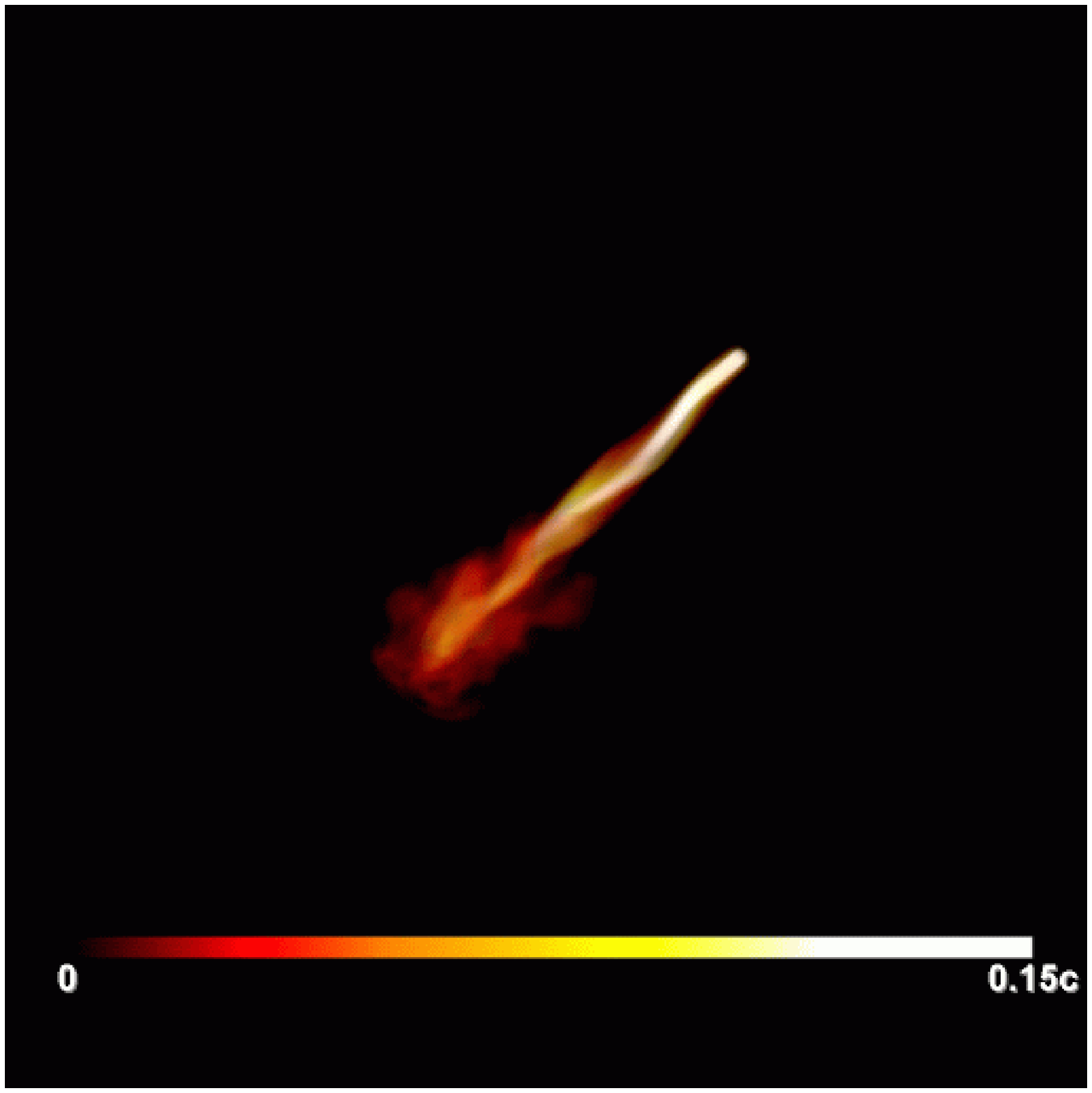}
\caption{As in Figure 1, for the LU model.}
\end{figure}

\begin{figure}
\epsscale{0.5}
\plotone{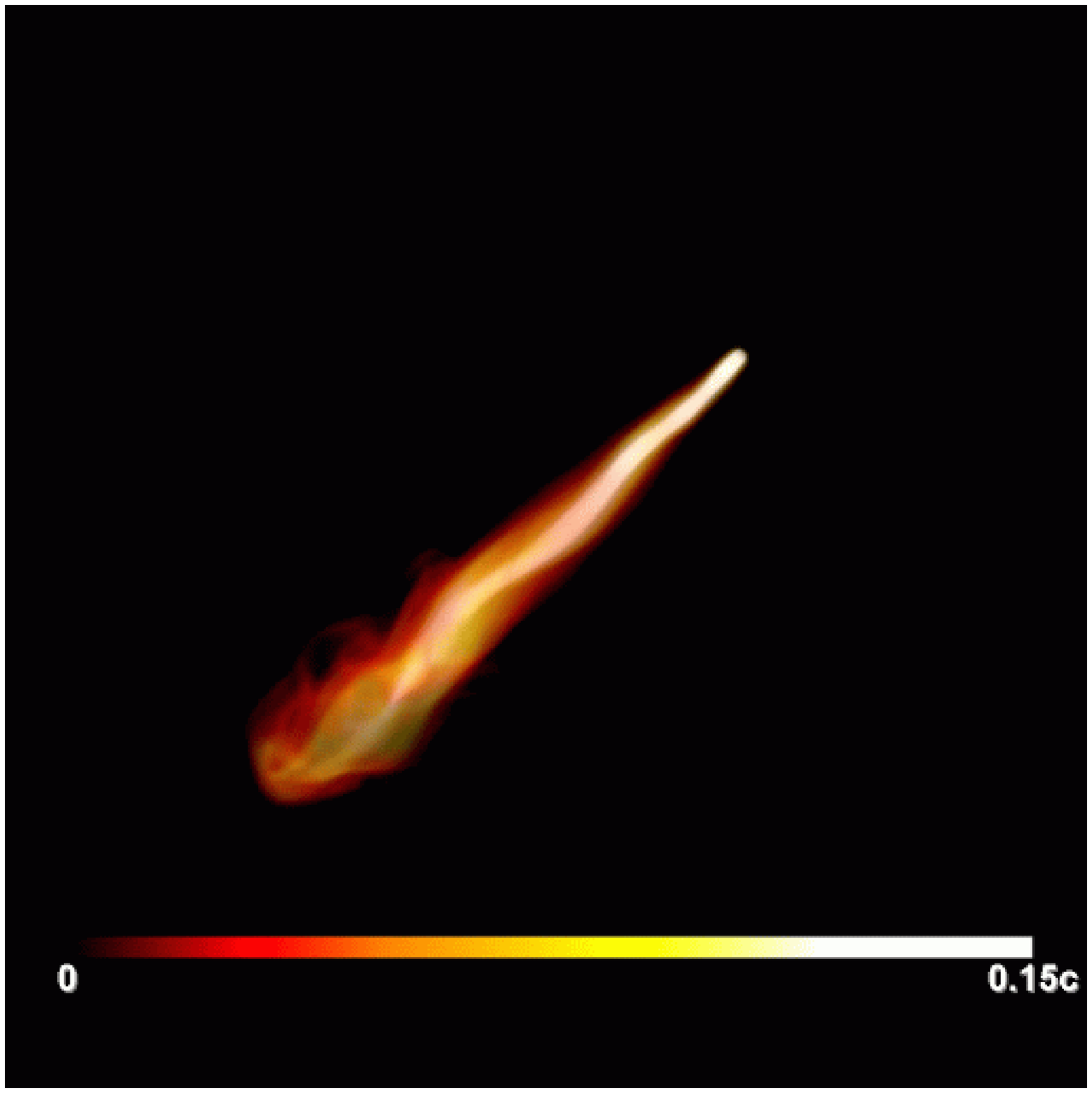}
\caption{As in Figure 1, for the LK model.}
\end{figure}

\begin{figure}
\epsscale{0.5}
\plotone{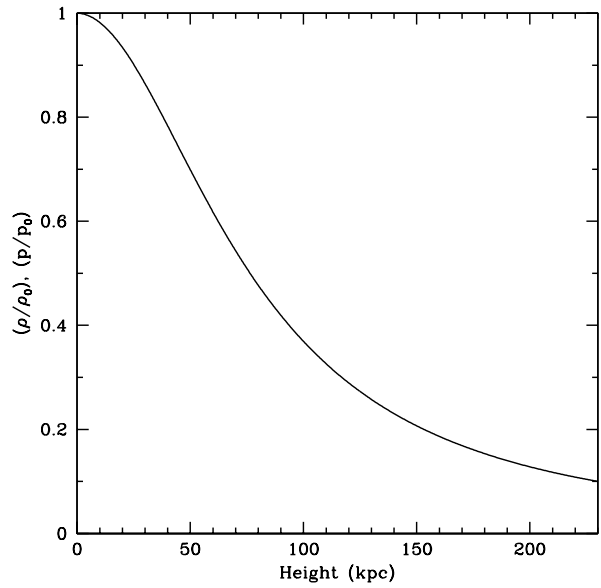}
\caption{The initial King-type density and pressure profiles, used in the HK and LK models.  Values of $\rho_0$ and $P_0$ are given in Table 1.}
\end{figure}

\begin{figure}
\epsscale{1}
\plotone{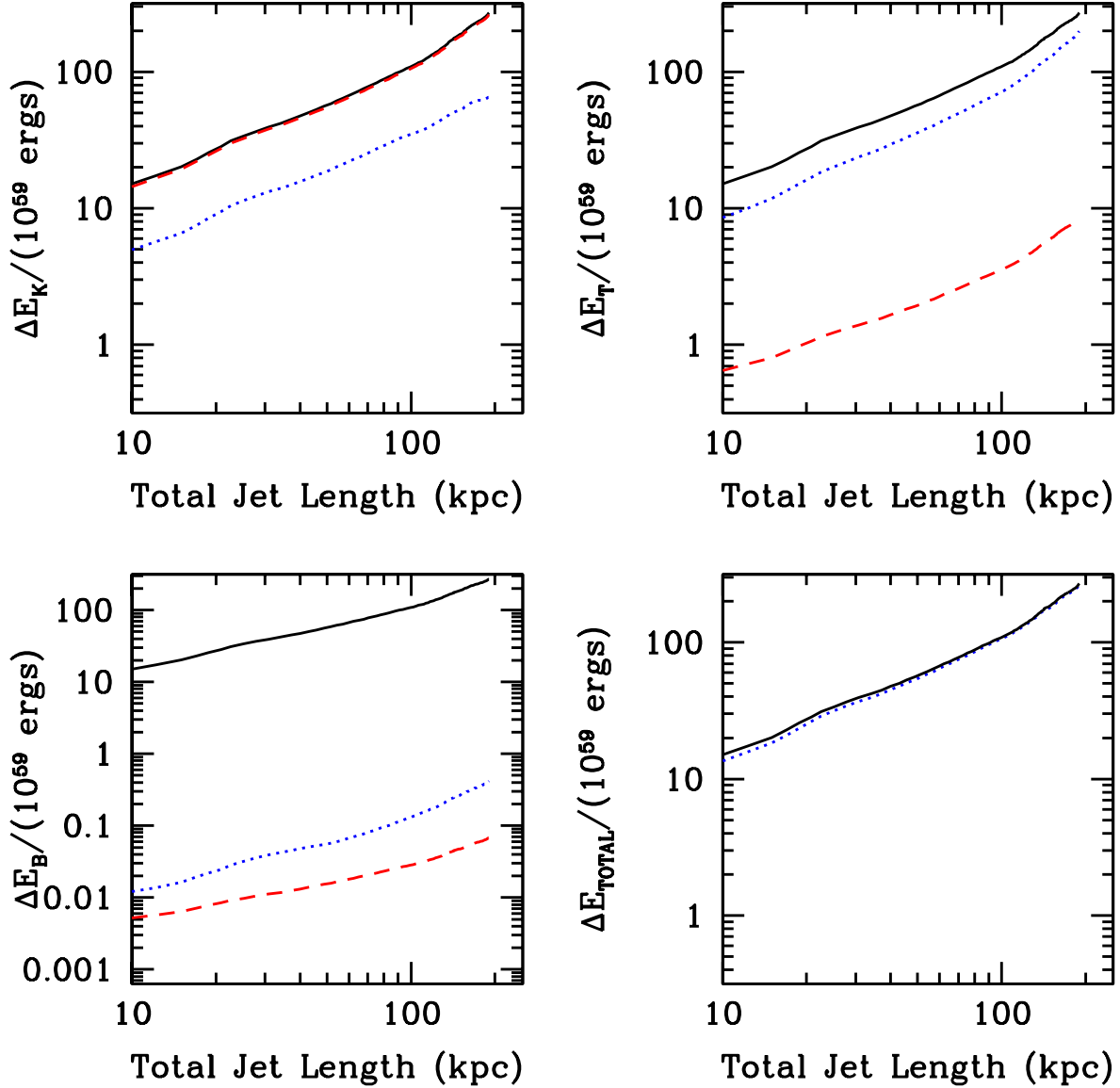}
\caption{Energy flow in the HU-r model.  Upper-left: A comparison of the known added kinetic energy ({\em dashed line}) to the measured change (relative to the initial value) in kinetic energy on the grid ({\em dotted line}) as they vary with the total jet length.  The total (kinetic + thermal + magnetic) inflow energy is shown ({\em solid line}) as a reference.  Upper-right: Same for the thermal energy.  Lower-left: Same for the magnetic energy.  Lower-right: Same for the total energy.}
\end{figure}

\begin{figure}
\plotone{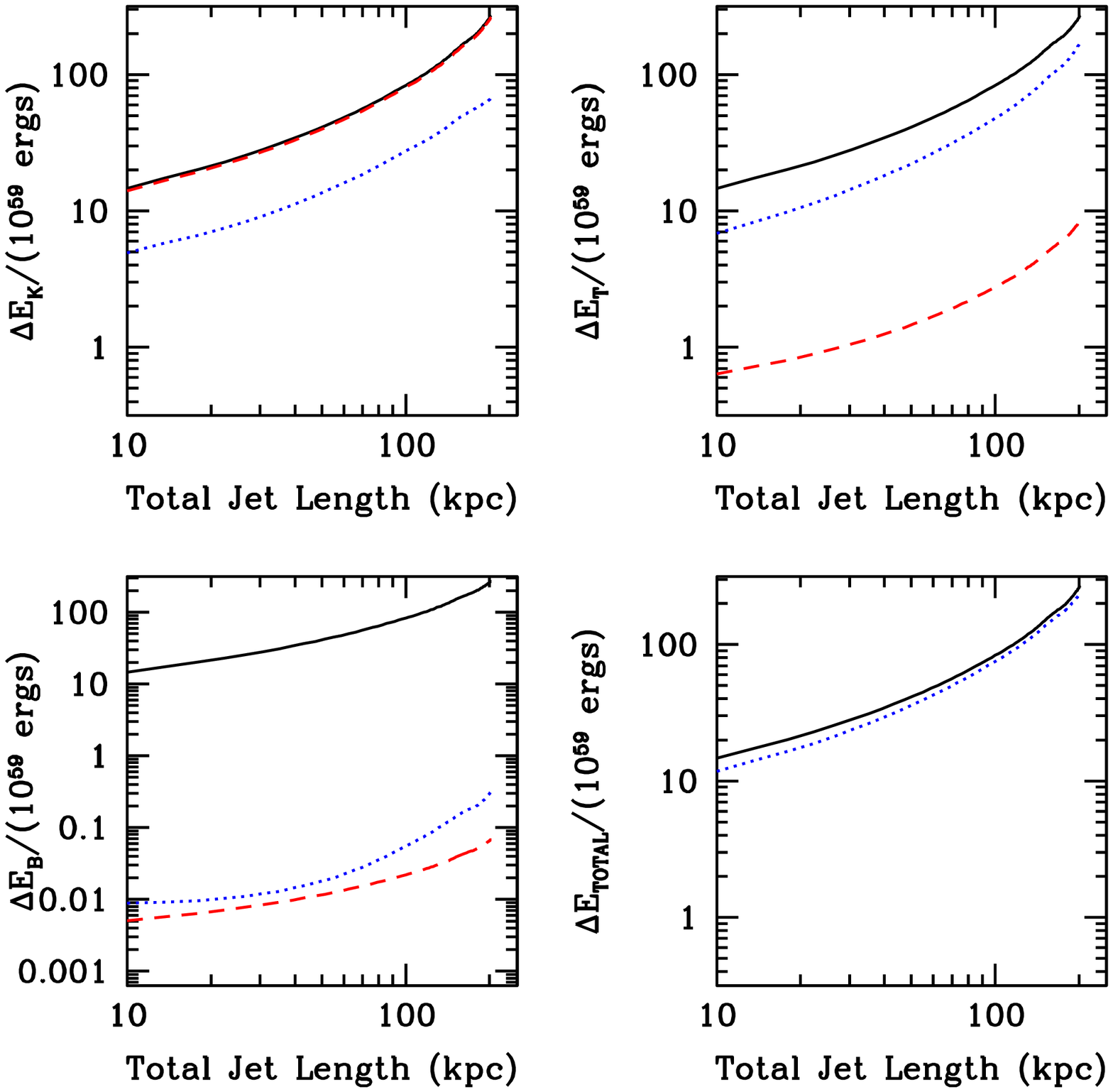}
\caption{As in Figure 7, for the HU model.}
\end{figure}

\begin{figure}
\plotone{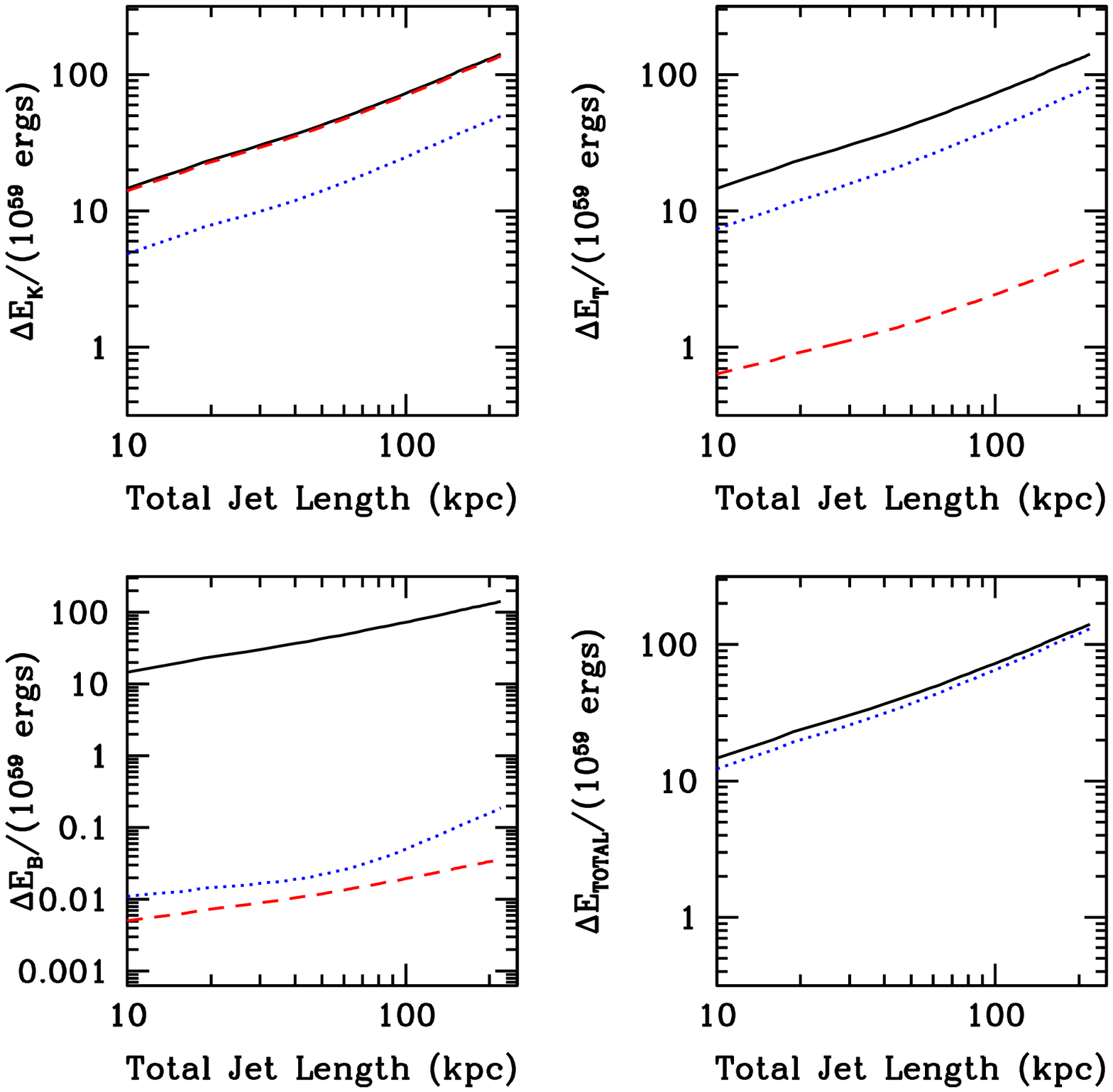}
\caption{As in Figure 7, for the HK model.}
\end{figure}

\begin{figure}
\plotone{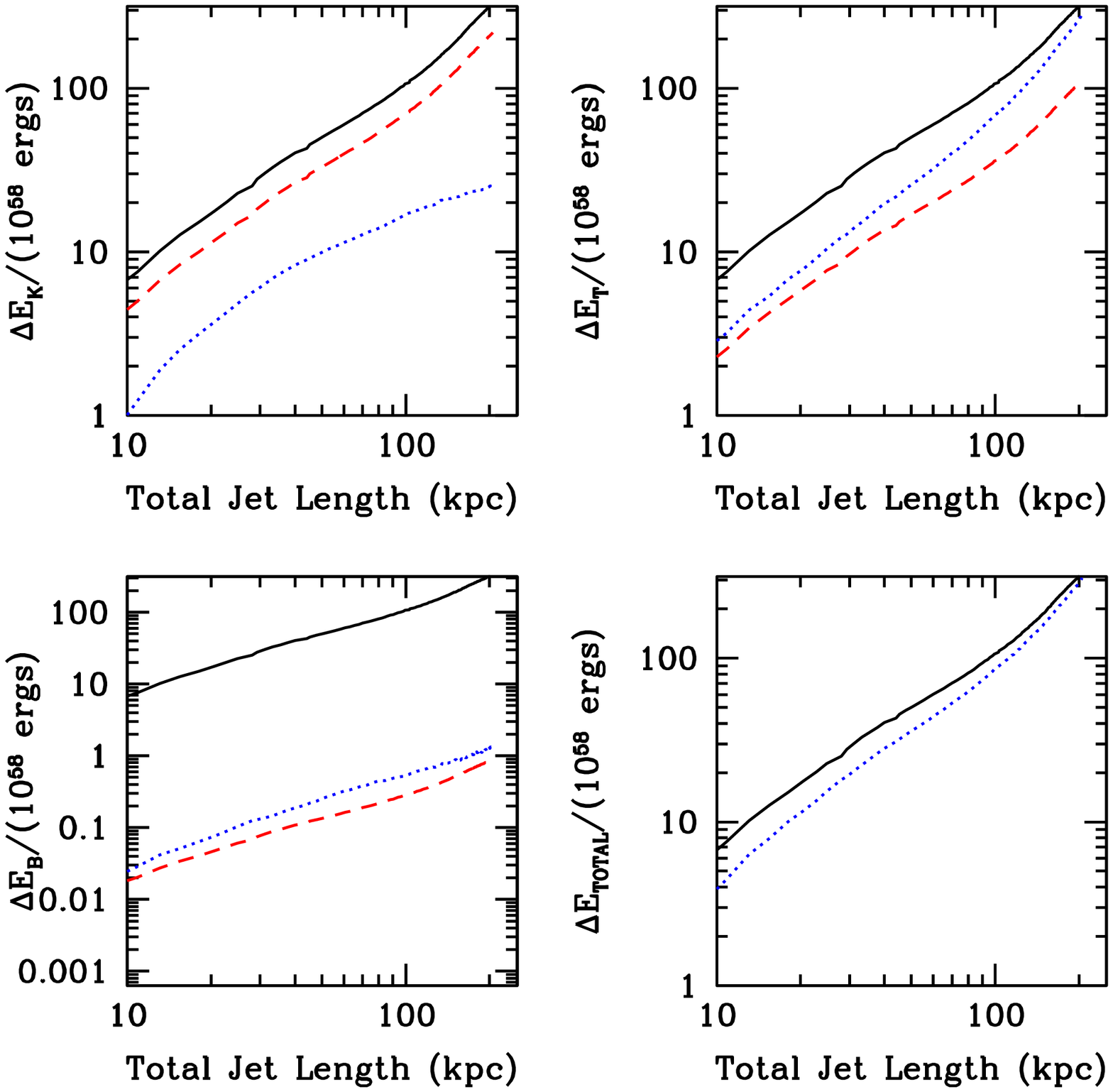}
\caption{As in Figure 7, for the LU model.}
\end{figure}

\begin{figure}
\plotone{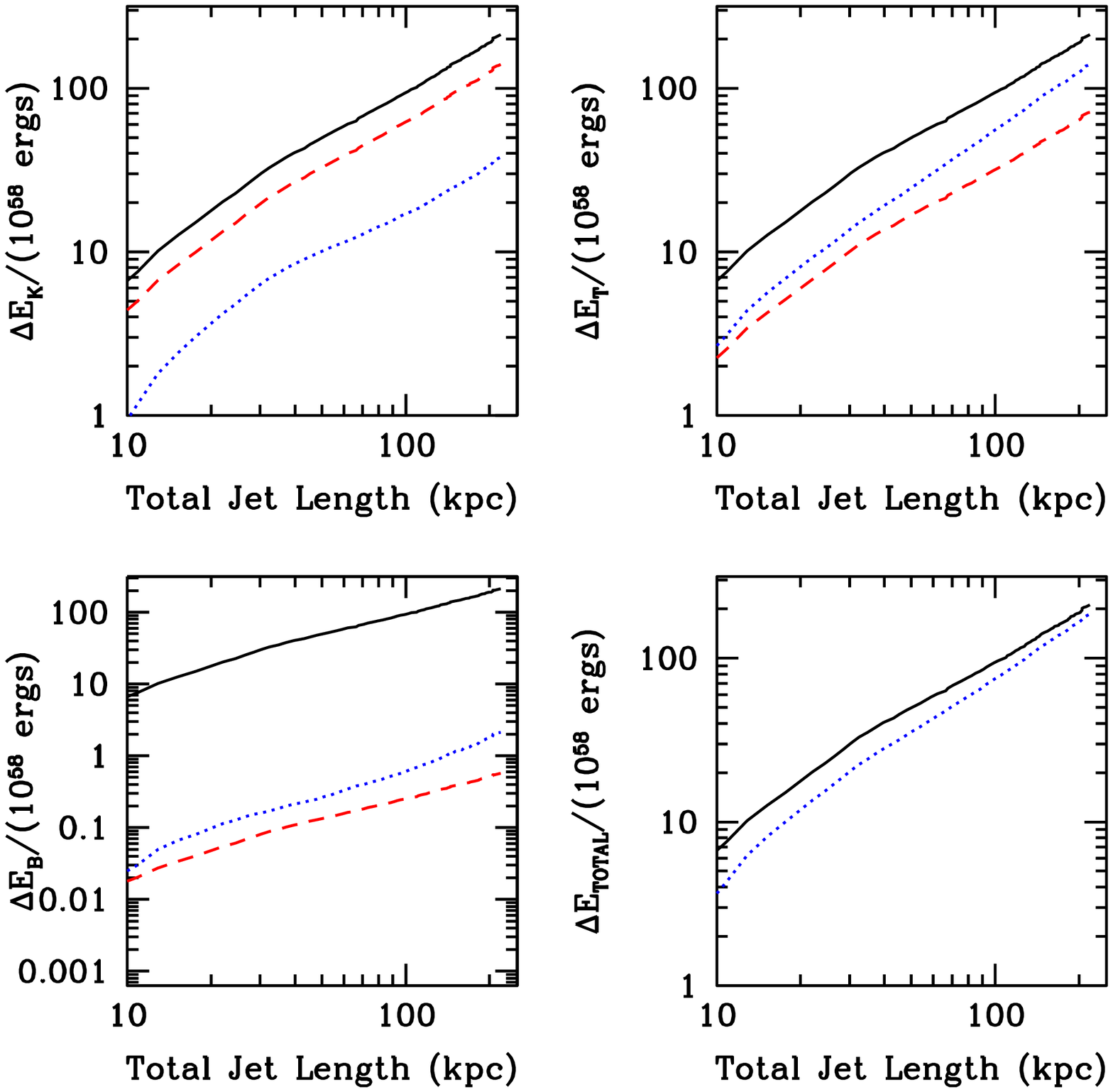}
\caption{As in Figure 7, for the LK model.}
\end{figure}

\begin{figure}
\plotone{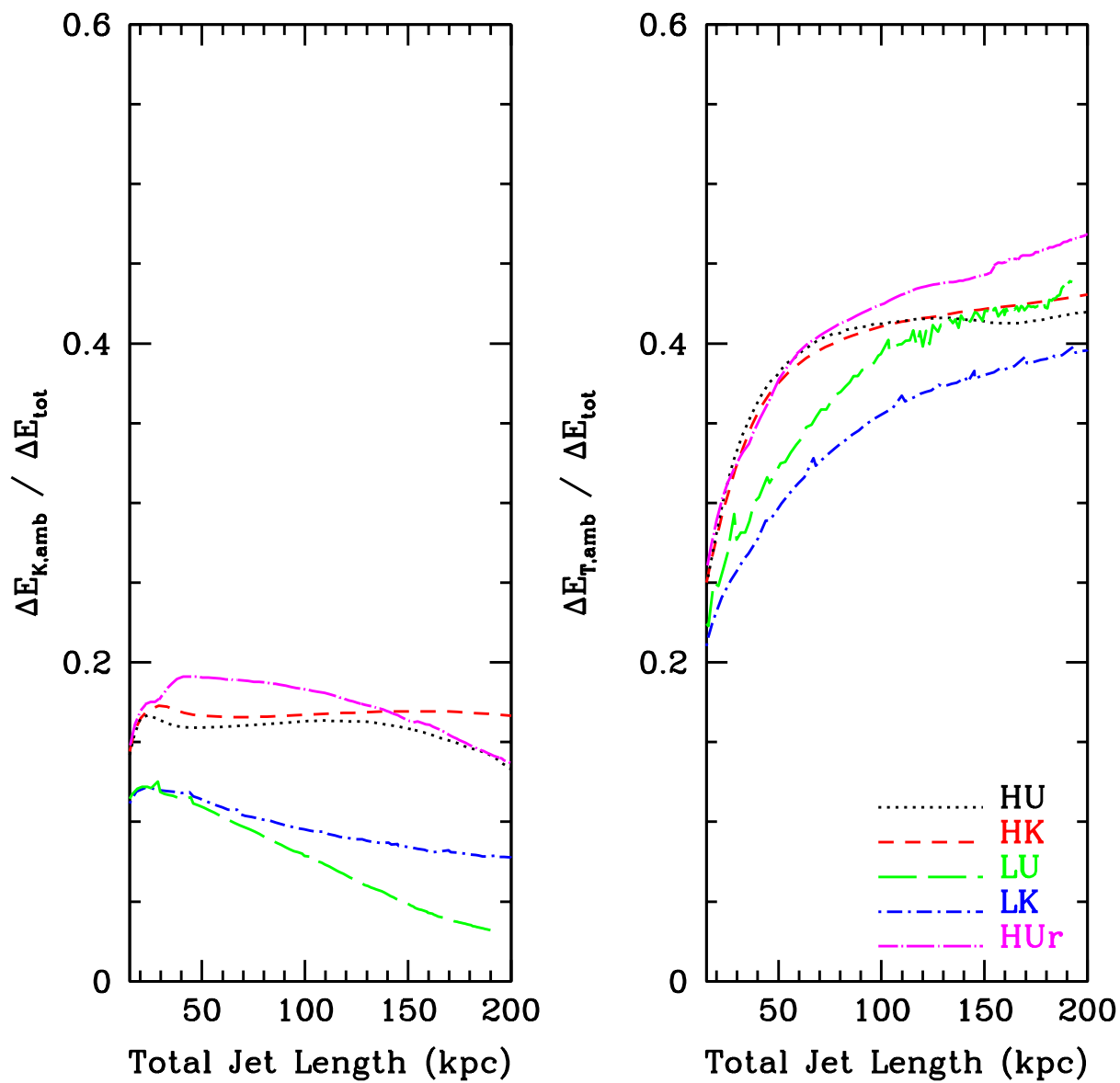}
\caption{Energy deposited in the ambient medium by the jets, as a fraction of total energy added to the grid.  The normalized kinetic energy added
to the ambient medium as a function of total jet length appears on the left while the thermal energy added to the ambient medium is shown on the right.}
\end{figure}

\begin{figure}
\epsscale{0.5}
\plotone{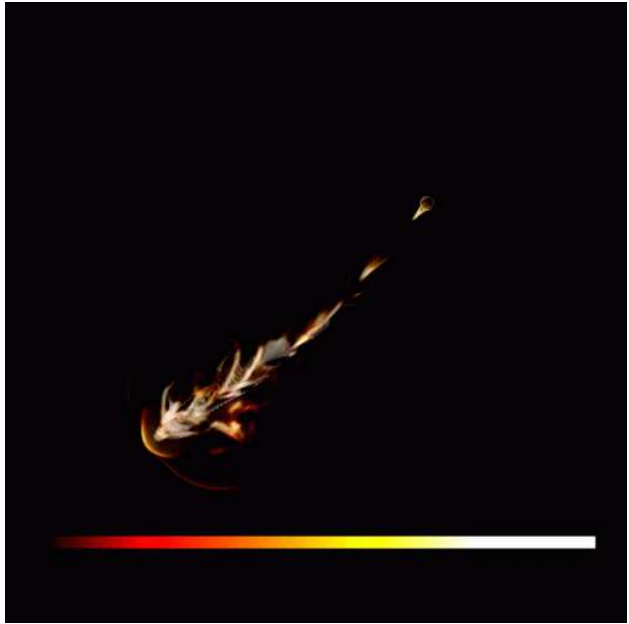}
\caption{Volume rendering of compression rate for the HU model, after the jet has propagated $\sim 90\%$ of the total grid length.  The colorbar runs from low compression rates (dim, dark) to high compression rates (bright, light).  An animation of this quantity as seen from several different angles appears at: http://www.msi.umn.edu/Projects/twj/newsite/projects/radiojets/movies/ }
\end{figure}

\begin{figure}
\plotone{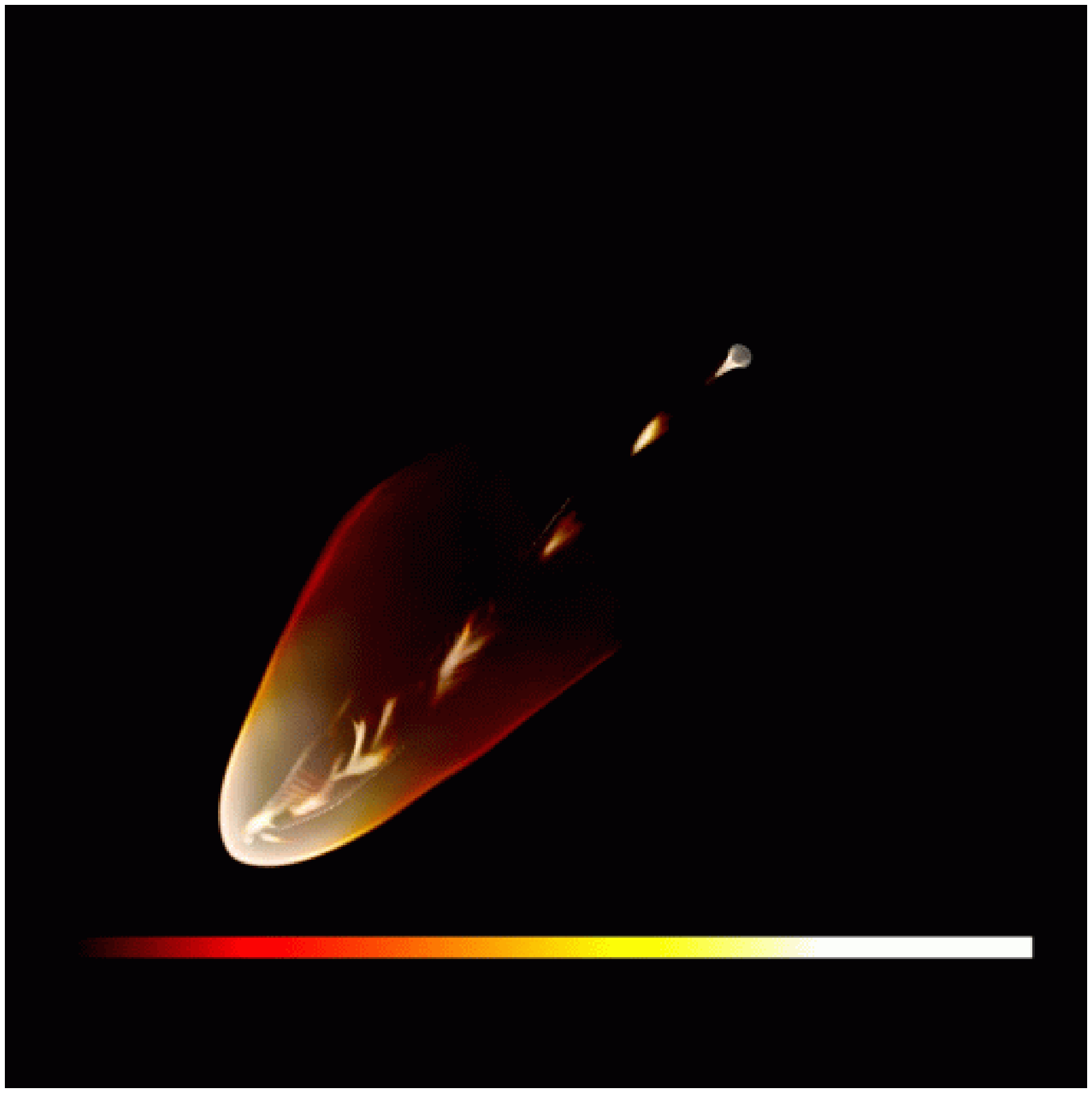}
\caption{As in Figure 13, for the HK model.}
\end{figure}

\begin{figure}
\plotone{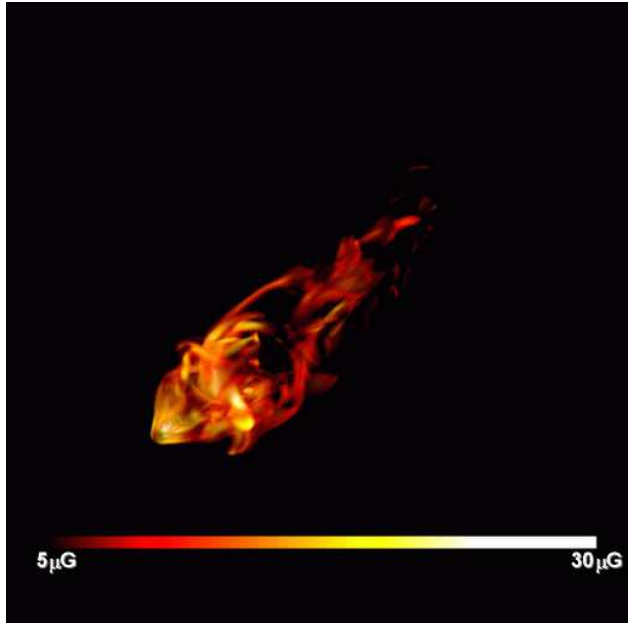}
\caption{Volume renderings of the log of magnetic field strength for the HU model, after the jet has propagated $\sim 90\%$ of the total grid length.  In this case, the bow shock has been removed to enhance visualization of the jet and cocoon.  An animation of this quantity as seen from several different angles appears at: http://www.msi.umn.edu/Projects/twj/newsite/projects/radiojets/movies/ }
\end{figure}

\begin{figure}
\plotone{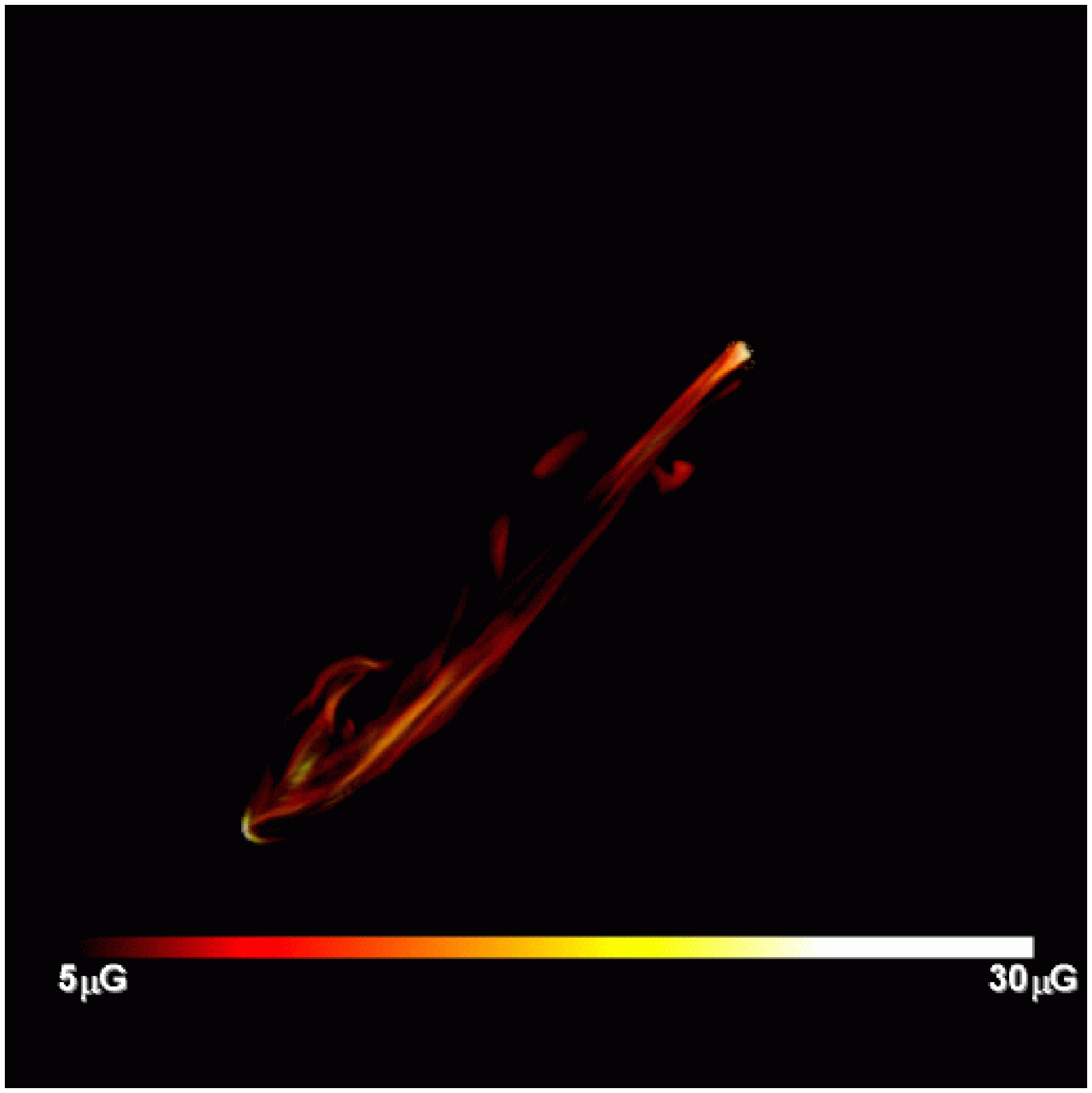}
\caption{As in Figure 15, for the HK model.}
\end{figure}

\begin{figure}
\epsscale{1}
\plotone{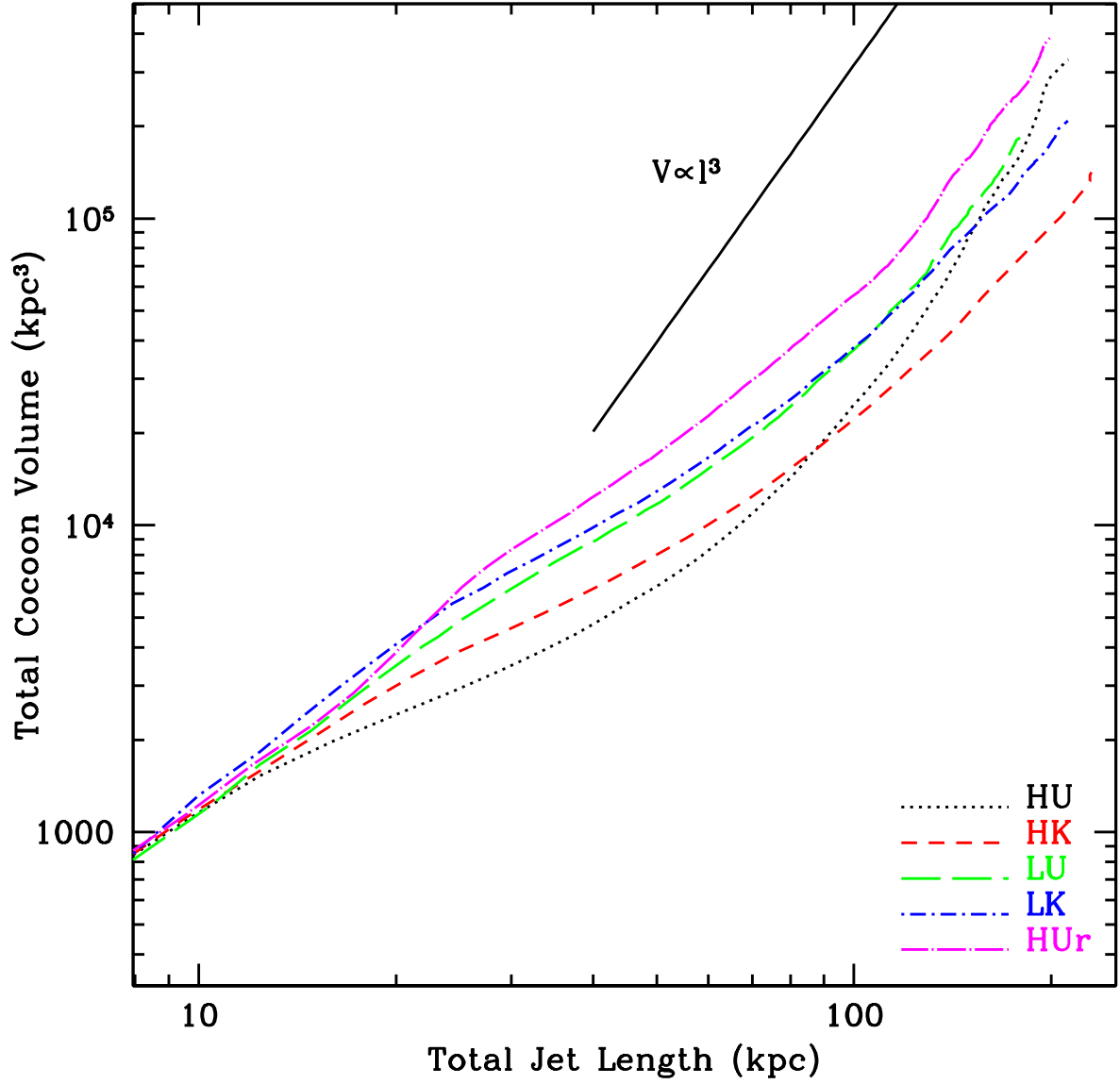}
\caption{Cocoon volumes are plotted as a function of total jet length for all models.  The solid line represents the slope for the simple scaling-law expectation.}
\end{figure}

\begin{figure}
\plotone{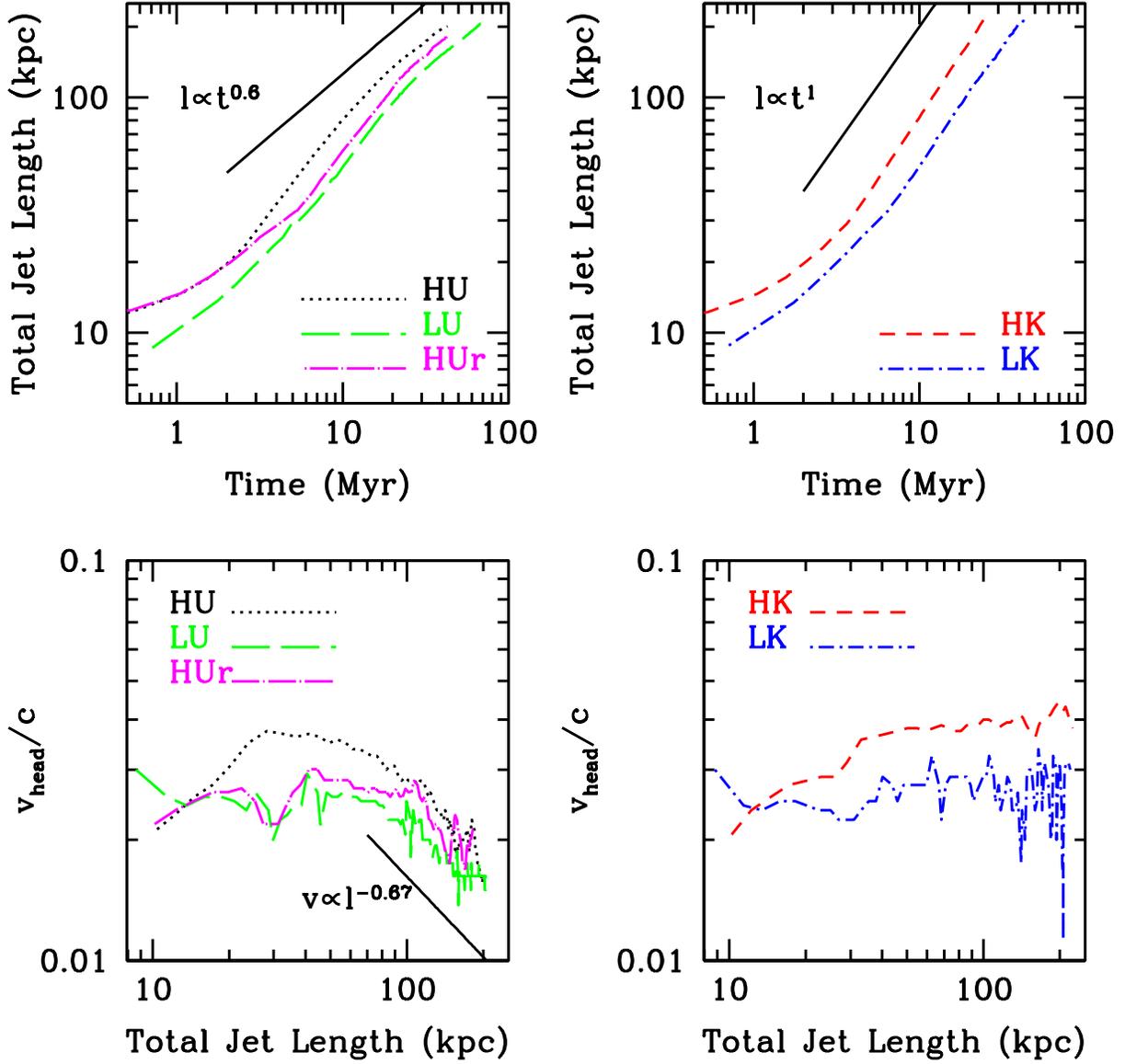}
\caption{Jet evolution: Jet length as a function of time for the uniform models (upper left) and the King models (upper right).  Jet head speed as a function of jet length for the uniform models (lower left) and the King models (lower right).  The solid lines represent the slopes for the simple scaling-law expectations.}
\end{figure}

\clearpage

\end{document}